# Laws of Energy Gradient for Instabilities

## Hua-Shu Dou

Temasek Laboratories, National Wind Tunnel Building,
National University of Singapore,
Singapore 117508, SINGAPORE
Email: tsldh@nus.edu.sg; huashudou@yahoo.com

**Abstract**

Transition to turbulence is due to the instability of a laminar flow subject to a disturbance. This complicated problem can be explained using a new proposed energy gradient theory in our previous study. This theory is extended to the instability of fluid material systems in this study. The instability of fluid material systems may lead to the evolution of natural environments and the occurrence of catastrophic events in the world. To better describe these phenomena and to understand the physical mechanism behind them are very important. In order to more generally describe the instability of fluid material systems, *laws of energy gradient* are summarized for static and motion systems, respectively. These laws could be applicable to various flow problems and material systems. Examples are shown that many events in the world could be explained using these laws.

## Introduction

Turbulence is a very complex phenomenon which has a history of more than 120 years. The mechanism of turbulence generation is still unknown so far. Earlier researches include theory, experiments and semi-empirical theory. In recent 30 years, numerical simulation for this topic has obtained large advance with the aid of the computer techniques. However, the understanding of the flow physics of turbulence is still limited. Even some experts assert that our research on turbulence is still at the enfant stage after 100 years [1]. Although we noticed that it is still lacking of theoretical unity in many respects, Tatsumi [2] stated that it is the time for constructing a unified physical theory of turbulence from the enormous amount of information obtained during this century.

Newtonian mechanics founded the basis of modern science and technology. However, even though using that, many phenomena could not be explained in the macro-world so far. The most typical one is the generation of turbulence as stated. This is the most difficult problem in classical mechanics.

Researches have shown that turbulence generation is resulted from the instability of laminar flows. All of the theories of flow instability, including linear theory, energy method, weak nonlinear theory, and secondary instability theory, could not satisfactorily explain the problem of flow instability in parallel flows [3-8]. Recently, Dou [9] proposed a new theory of energy gradient for flow instability and turbulence transition. The energy gradient theory obtains consistent results for the subcritical transition of parallel flows. Dou showed that the energy gradient theory is a better method for the description of subcritical flow transition than the others.

Many phenomena occurring in the nature have similar behaviour as the generation of turbulence. Many material systems have the feature of flow like fluid. This is as described in [10]: everything flows. Therefore, these systems may be described using same theory as for turbulence.

In this paper, we show that energy gradient theory is suitable for stability of various fluid material systems in nature. This theory may be used for the prediction of catastrophic events in the nature. New laws are proposed for the Newtonian mechanics.

## Energy Gradient Theory

Dou (2004) [9] proposed an energy gradient theory with the aim to clarify the mechanism of transition from laminar to turbulence. It is thought that the gradient of total energy in the transverse direction of the main flow and the viscous friction in the streamwise direction dominate the instability phenomena and hence the flow transition. The energy gradient in the transverse direction has the potential to amplify a velocity disturbance, while the viscous friction loss in the streamwise direction can resist and absorb this disturbance energy. The transition to turbulence depends on the relative magnitude of the two roles of energy gradient amplifying and viscous friction damping to the initial disturbance. Based on such, a new dimensionless parameter, $K$ (the ratio of the energy gradient in the transverse direction to that in the streamwise direction), can be written as,

$$K = \frac{\partial E / \partial n}{\partial E / \partial s} \quad (1)$$

Here, $E = p + \frac{1}{2}\rho V^2 + \rho g \xi$ is the total energy for incompressible flows with $\xi$ as the coordinate perpendicular to the ground, $n$ denotes the direction normal to the streamwise direction and $s$ denotes the streamwise direction. $\rho$ is the fluid density, g the gravity acceleration, V the velocity, and p the hydrodynamic pressure. The parameter $K$ in Eq.(1) is a field variable. The occurrence of instability depends upon the magnitude of this dimensionless parameter $K$ and the critical condition is determined by the maximum value of $K$ in the flow. For a given flow geometry and fluid properties, when the maximum of $K$ in the flow field is larger than a critical value $Kc$, it is expected that instability would occur for certain initial disturbance [9]. The analysis showed that the transition to turbulence is due to the energy gradient and the disturbance amplification, rather than a linear eigenvalue instability type [11, 12]. Both Grossmann [11] and Trefethen et al.'s [12] commented that the nature of the onset-of-turbulence mechanism in parallel shear flows must be different from an eigenvalue instability of linear equations of small disturbance. Dou (2004) demonstrated that the criterion has obtained excellent agreement with the experimental data for plane Poiseuille flow and pipe Poiseuille flow as well as plane Couette flow, see Table 1 [9]. It can be found that the turbulence transition takes place at a critical value of Kc of about 385 for both plane Poiseuille flow and pipe Poiseuille flow, and about 370 for plane Couette flow, and they obtained a consistent value. This result proved that the flow instability is resulted from the action of energy gradients, but not the kind of eigenvalue instability of linear equations. The comparison of theory with experiments is shown in Table 1 for wall bounded parallel flows (Fig.1).

The proposed principle can be used to both pressure and shear driven flows. If we assume that there is no energy input (such as shear) to the system or energy output from the system, this criterion can predict that the viscous flow with inflectional velocity is unstable. This is because if there is an inflection point in the velocity profile, the value of maximum of K in the flow will be infinite. Therefore, the flow is unstable at the inflection point. Following this principle, it is proved that viscous parallel flow with inflectional velocity profile is sufficient for flow instability for both two-dimensional and axisymmetrical flows [14].

| Flow type | Re expression | Linear theory, $Re_c$ | Exp $Re_c$ | $K_{max}$ at Exp $Re_c$ |
|---|---|---|---|---|
| Pipe Poiseuille | $Re = \rho UD/\mu$ | $\infty$ | 2000 | 385 |
| Plane Poiseuille | $Re = \rho UL/\mu$ | 7696 | 1350 | 389 |
|  | $Re = \rho u_0 h/\mu$ | 5772 | 1012 | 389 |
| Plane Couette | $Re = \rho Uh/\mu$ | $\infty$ | 370 | 370 |

Table 1 Comparison of the critical Reynolds number and the energy gradient parameter Kmax for plane Poiseuille flow and pipe Poiseuille flow [9] as well as for plane Couette flow [13].

Although laminar-to-turbulent transition can occur through several mechanisms, such as in linear instability, bypass transition (skip linear instability), Gortler instability (flow on concave surface), and cross-flow instability (flow over swept wing) [15], all of these instabilities can be included in the frame of instabilities resulted by energy gradient.

The instability mechanism can also get some hint from solid mechanics. As is well known, the damage of a metal component generally starts from some area such as manufacturing fault, crack, stress concentration, or fatigue position, etc. In fluid mechanics, the breaking down of a steady flow should also start from some most dangerous position first. For example, for the flow around an airfoil at a large attack angle, the flow instability first starts from the rear part on the suction side where the pressure gradient is large. For the flow around a cylinder, it is known that the flow instability begins first from the two inflection points near the rear stagnation point.

## Laws of Energy Gradient For Flow Instability

The Newtonian mechanics as a basic subject has been developed for more than 300 years since this subject was established (1664-1684). The Newton's three laws (Law of inertia, law of acceleration, and law of action and reaction) established the foundation of modern mechanics and becomes the backbone of modern science and technology [16].

In these laws, it is not shown why a fluid material system becomes of motion from static (or why a phase change occurs), and why the flow of materials becomes unstable from stable. Now, we describe the laws for the stability of fluid material systems using the energy gradient theory.

**First law of energy gradient (static system):**

*If a material system is static, when the energy gradient in some direction is larger than a critical value, the system will become unstable and the phase change or flow would occur.*

The unstable condition can be expressed as

$$\frac{\partial E}{\partial x} > C, \qquad (2)$$

where C is the critical value which is related to the material properties (friction factor, density, etc.) of the material and the geometry of the problem. This law may be restated as: when the energy gradient in some direction is larger than the work which should be done to overcome the resistance for the material moving, the material will be unstable and the material may flow or the phase change may occur. Thus, this law may be expressed as that the unstable condition of a system is,

$$D = \frac{\partial E/\partial x}{\partial A/\partial x} > 1, \qquad (3)$$

where $\partial A/\partial x = C$ is the work needed to make the material moving for unit volume media. Actually, this law is consistent with the first law and second law of Newton. Newton's first and second laws describe the system from the roles of forces, while the first law of energy gradient describes the system from the viewpoint of energy field.

**Second law of energy gradient (moving system):**

*If a material system flows, when the ratio of the energy gradient in the transverse direction and that in the streamwise direction is larger than a critical value, the system will become unstable.*

The unstable condition for a moving system can be expressed as

$$K = \frac{\partial E/\partial y}{\partial E/\partial x} > K_c, \qquad (4)$$

where $K_c$ is the critical parameter which is related to the material properties and the flow geometry. The instability of plane Poiseuille flow, pipe Poiseuille flow, and plane Couette flow are typical examples of instability of moving systems.

These laws could be thought as that they are supplements to Newton's three laws. *Any fluid system violating these laws will become unstable.* The energy gradient laws enriched the system of classical mechanics. They can deal with problems which could not be resolved by Newton's laws. This will be demonstrated in later section for use in various instability problems. From these laws, it can be found that the reason of a material system moving in some cases is due to energy gradient, and not simply to the role of forces. For example, fluid cell formation in Rayleigh-Benard problem is not resulted in by forces, but by energy gradient (we will show the detail later). Energy gradient is the power of motion of material when there is no external force acting on it. Further, it is found that the motion of media from one state to another state is due to the role of energy gradient. The instability of the media state is caused by energy gradient, and a new state is then formed through the motion of the media under the energy gradient.

## Examples of Material System Instabilities

We take x is horizontal and y is upward. The total energy per unit volume fluid for incompressible fluids can be written as

$$E = p + \frac{1}{2}\rho V^2 + \rho gy. \qquad (5)$$

**Static fluid in a container:** (a) If the fluid in the container is single fluid (Fig.2a), there is $\frac{\partial E}{\partial x} = 0$ and $\frac{\partial E}{\partial y} = 0$ in everywhere of the domain. Thus, the energy of fluid in the domain is uniform. Because there is no energy gradient in the domain, the fluid is stable. (b) If one fluid ($\rho_1$) is over another fluid ($\rho_2$) in the container, the energy gradient will be formed at the interface (fig.2b). When $\rho_1 < \rho_2$, there are $\frac{\partial E}{\partial x} = 0$ and $\frac{\partial E}{\partial y} < 0$ at the interface, the fluid state is stable. When $\rho_1 > \rho_2$, there are $\frac{\partial E}{\partial x} = 0$ and $\frac{\partial E}{\partial y} > 0$ at the interface, the fluid state is unstable (Rayleigh-Taylor instability).

**Uniform flow will be stable:** For uniform flow, the velocity is a constant in the domain (Fig.2c). If the fluid gravity is neglected, there are $\frac{\partial E}{\partial x} = 0$ and $\frac{\partial E}{\partial y} = 0$ in everywhere of the flow field, and thus the flow is stable.

**Kelvin-Helmholtz instability:** Free shear flow displays an instability called Kelvin-Helmholtz instability (Fig.2d). There are $\frac{\partial E}{\partial x} = \nu \nabla^2 \mathbf{V}$ and $\frac{\partial E}{\partial y} = V \frac{\partial V}{\partial y}$ in the flow field if the fluid gravity is neglected. The flow stability depends on the magnitude of K calculated by Eq.(1). Because an inflection point exists in the interface of free shear flow, we have $K = \infty$ at the inflection point from Eq.(1). Therefore, free shear flow is always unstable.

**Rayleigh-Bernard problem:** When the fluid is placed on a horizontal plate and it is heated from below. Then, the fluid will become unstable, and fluid cells of vortices will be formed (Fig.2e). This kind of pattern of roll formation can be explained using the energy gradient theory. If we treat the fluid as incompressible, $E = p + \frac{1}{2}\rho V^2 + \rho g y$, then $\frac{\partial E}{\partial x} = 0$ and $\frac{\partial E}{\partial y} = gy \frac{\partial \rho}{\partial y} > 0$ in the fluid domain because the fluid density in the bottom becomes low. If $\frac{\partial E}{\partial y}$ is larger than a critical value, the flow will becomes unstable. It is noticed that there is no role of forces, but there is an energy gradient in the vertical direction.

**Stratified flows:** When two layers of fluids with different densities flow along one direction, instability due to stratified density will occur. This is because an energy gradient is formed at the interface (Fig.2f). The formulation is similar to that as for Fig.2e.

**Wall bounded Parallel flows:** Plane Posiseuille flow, pipe Poiseuille flow, and plane Couette flow are all the examples of instabilities of parallel flows resulted by energy gradients, as shown in Table 1 and Fig.1. These flows have been studied in [9] and [13].

**Inflectional instability:** Flows behind a cylinder display instability at higher Reynolds number. It is known that this instability is caused by the inflection points near the rear stagnation points. Kelvin-Helmholtz instability is another example of inflectional instability produced by jet-wake (Fig.2d) as stated before.

**Granular material instability (Fig.3):** (a) Avalanche of piled sands or phase change (Fig.3a): A sand pile is placed on an inclined plate. When the inclined angle exceeds a critical value, the avalanche of sand pile will occur due to the energy gradient influence (gravitational energy).

**Migration of stone in the bottom of a river bed (Fig.3b):** Stones on the bottom of a river is moved by water flow. This process could be explained using the energy gradient theory (pressure energy).

**Migration of sands in desert (Fig.3c):** Sand piles in desert moves due to the role of winding around them. This movement could be explained using the energy gradient theory (pressure energy).

The experimental data for material system is very limited and the critical value also depends on the material property. The analyses for instabilities of granular materials in literature are mostly qualitative. Here, we summarize the comparison of laws of energy gradient with experimental observations for a few cases in Table 2.

| Fluid system | Experiments observations | Laws of Energy gradient |
|---|---|---|
| Single fluid, Fig.2a | Stable | Stable |
| Two fluids, $\rho_1 > \rho_2$, Fig.2b | Unstable | Unstable |
| Uniform flow, Fig.2c | Rec=∞, stable | Rec=∞, stable |
| Free shear flow, Fig.2d | Rec=0, unstable | Rec=0, unstable |

Table 2  Comparison of theory with experiments.

Energy gradient is not only the mechanism of turbulence generation, but also the rule of generation of many unstable phenomena in nature. For example, evolution of cosmic stars, motion of mantle of earth, earthquake occurrence, land shift, mountain coast, snow avalanche, dam breaking, change of atmosphere, etc, all can be explained and described using the energy gradient laws. The breakdown of the process of these mechanical systems can be universally described in detail using this theory. Sornette (2002) described a unifying approach for modelling and predicting these catastrophic events or ''ruptures,'' that is, sudden transitions from a quiescent state to a crisis [17]. He stated that such ruptures involve interactions between structures at many different scales. He believes that it is possible to develop universal theory and tool for predicting these catastrophic events. The energy gradient laws can be used to predict the flow-system breakdown and the catastrophic events too [18]. The energy gradient theory [9] describes a mechanism of failure of mechanical system in a simple way. It will demonstrate its powerfulness in the application of catastrophic predictions in near future.

**Conclusion**

Energy gradient theory was proposed for laminar-turbulence instability [9]. In present work, it is shown that this theory is also suitable for instabilities of various material systems in nature. Laws are proposed for describing these instabilities. These laws can be considered as a supplement to the Newtonian mechanics, which enriched the theoretical system of the classical mechanics. Newton's first and second laws described the rules of body motion from the equilibrium of forces. Laws of energy gradient describe the rules of fluid material stability from the viewpoint of energy field. These laws are more general because it described the stability of moving fluids and those under no force. Listed examples have been shown that the validity of the proposed laws. On the other hand, using laws of energy gradient we can control the evolution of fluid material systems.

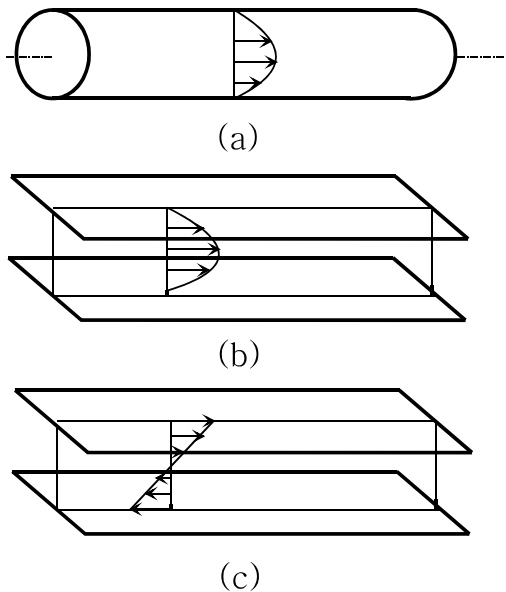

Fig.1 Wall bounded parallel flows. (a) Pipe Poiseuille flow; (b) Plane Poiseuille flow; (c) Plane Couette flow.

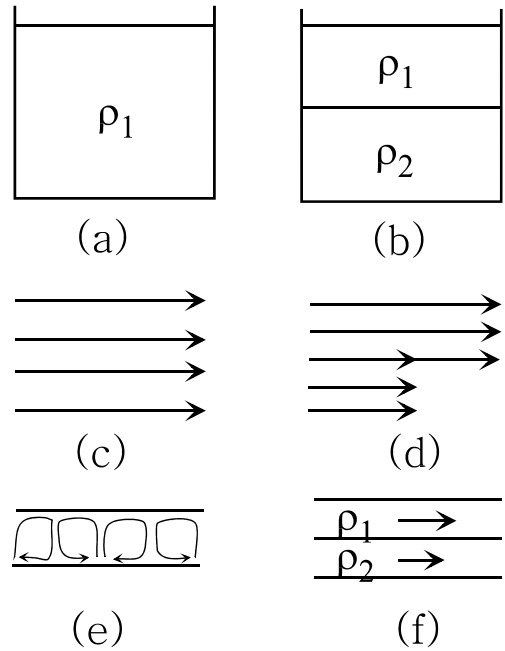

Fig.2 Flow instabilities resulted by energy gradient. (a) Single fluid in a container; (b) Two fluids in a container; (c) Uniform flow; (d) Free shear flow; (e) Rayleigh-Bernard flow; (f) Stratified flow by density.

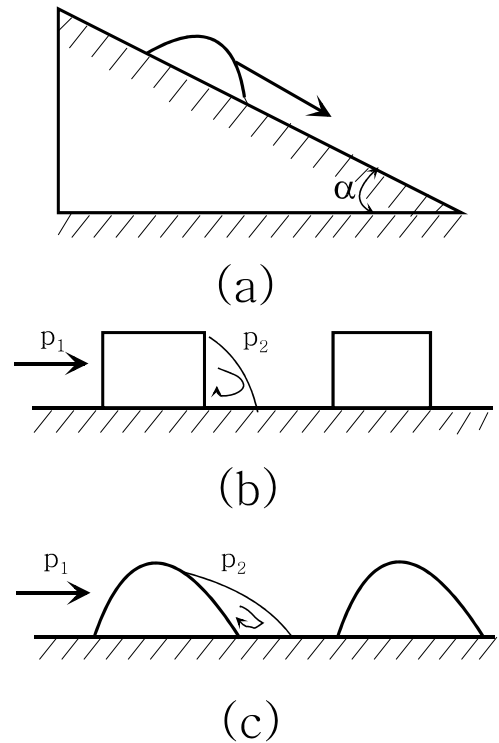

Fig.3 Materials moving under energy gradient. (a) Granular material avalanching on an inclined plate; (b) Stone moving in a river bed. (c) Sand pile moving in a desert.